\documentclass[12pt]{article}
\usepackage{latexsym}
\usepackage{amsfonts}
\usepackage{mathrsfs}
\usepackage[cp1250]{inputenc}

\title{Nonlinear realizations and the orbit method}
\author{Joanna Gonera\thanks{jgonera@uni.lodz.pl},\\ 
Department of Theoretical Physics and Computer Science \\
University of {\L}\'od\'z \\
Pomorska 149/153, 90 - 236 {\L}\'od\'z/Poland.}
\date{}

\begin{document}
\maketitle
\begin{abstract}
Given a symmetry group one can construct the invariant dynamics using the technique of nonlinear 
realizations or the orbit method. The relationship between these methods is discussed. Few examples
are presented.
\end{abstract}
\newpage
\section{Introducion}

It is very common in physics that the symmetry principles provide the starting point for constructing the relevant
dynamics, both on classical and quantum levels. Among numerous examples we quote only a few:
Poincare \cite{n1} and Galilean \cite{n2} symmetries, Newton-Hooke group \cite{n3}, relativistic \cite{n4} and nonrelativistic \cite{n5}
conformal groups etc.

The natural question arises whether there exist methods which allow to find and, if possible, to classify the dynamical 
systems exhibiting a given in advance symmetry. 
At least in the case of Hamiltonian systems with finite number of degrees of freedom the corresponding algorithm does exist and is called
the orbit method \cite{n6}-\cite{n11}. It allows, in particular, to classify all symplectic manifolds on which a given group $G$ acts 
transitively and preserves symplectic structure. 

An alternative method relies on the theory of nonlinear realizations of groups \cite{n12}. It allows to construct various invariant dynamical 
systems in terms of geometry of group manifolds. It was successfully applied to the case of conformal $SL(2,\mathbb R)$ symmetry \cite{n13} 
in the interesting paper by Ivanov et al. \cite{n14} (for supersymmetric version see \cite{n15}). Following their method Fedoruk at al. \cite{n16}
considered dynamical realizations of $N=2$ conformal Galilei algebras; this was further generalized \cite{n17} to arbitrary $N$.

The method of nonlinear realizations has many advantages. It allows to write out invariant equations of first order once a subgroup of 
symmetry group is selected and the relevant Cartan-Maurer forms computed. On the other hand, it cannot be taken for granted that the
resulting equations can be put in Hamiltonian form; in fact, it is the case only provided the subgroup selected is the stability
subgroup of same point on coadjoint orbit.

In the present paper we analyze in some detail the relation between the method based on nonlinear realizations and the orbit method. In Sec.II we
show that the orbit method can be reformulated in terms of nonlinear action of group on coset space. In particular, the Hamiltonian equations of motion 
are formulated with the help of Cartan-Mauerer forms and the group action in both formalisms is described.\\
In Sec.III few examples are considered. First, the results of Ref. \cite{n14} are discussed within the general framework presented in Sec.II.
Then, two examples related to $N=2$ conformal Galilei group are described. While for the (centrally extended) $N$ odd conformal Galilean groups 
the situation seems to be clear \cite{n18}-\cite{n20}, the case of $N$ even is more complicated. The algebra does not allow central extension (except in
two dimensions) and the classification of coadjoint orbits is much more involved. Therefore, we restrict ourselves to two special cases.
 The first onereduces to the $SL(2,\mathbb R)$-invariant dynamics; the second describes nontrivial yet simple dynamics. 
Sec.IV is devoted to short conclusions. 
\newpage
\section{The orbit method and nonlinear realizations}

The orbit method \cite{n6}-\cite{n11} provides a powerful tool for constructing the Hamiltonian systems with dynamical symmetry.
In fact, it allows to classify all phase manifolds (i.e. symplectic ones) on which a given Lie group $G$ acts transitively as a group
of canonical transformations. To specify the dynamics the Hamiltonian is assumed to be an element of the Lie algebra of $G$ (or, more
generally, its universal enveloping algebra). The power of the method relies on the fact that the dynamical systems can be defined and
classified by making only general assumptions concerning the symmetry structure; no further a priori assumptions have to be made.

Let us discuss the main ingredients of the orbit method. Let $G$ be a Lie group with Lie algebra $\mathcal G$ spanned by the generators $X_i$
obeying the commutation rules
\\
\begin{equation}
\label{1}
\left[ X_i, X_j \right] = ic_{ij}\,^k X_k
\end{equation}
\\
Any element of $\mathcal G$ can be written as
\\
\begin{equation}
\label{2}
\xi = \xi ^i X_i
\end{equation}
\\
The adjoint action of $G$ is defined by
\\
\begin{equation}
\label{3}
Ad_g(X_i) = g X g^{-1} \equiv D^j\,_i(g) X_j
\end{equation} 
\\
In particular, for one-parameter subgroups $ g = \exp{\{it\xi\} }$ one obtains 
\\
\begin{equation}
\label{4}
Ad_g(X_i) = e^{it\xi } X_i e^{-it\xi } = X_i - t \xi ^kc_{ki}\,^j X_j + O(t^2)
\end{equation}
\\
or, equivalently
\\ 
\begin{equation}
\label{5}
D^j\,_i \left( e^{it\xi} \right) = \delta ^j\,_i - t\xi ^kc_{ki}\,^j + O(t^2) =  \delta ^j\,_i + t\xi ^kc_{ik}\,^j + O(t^2) 
\end{equation} 
\\
For a general element $\xi \in \mathcal G$ we get
\\
\begin{equation}
\label{6}
Ad_g \xi = {\xi '}^iX_i,\qquad {\xi '}^i = D^i\,_j(g)\xi ^j
\end{equation}
\\
The basis $\left\{ X^i\right\}$ in the dual space $\mathcal G^*$ is defined by pairing
\\ 
\begin{equation}
\left < X^i, X_j \right> = \delta ^i\,_j
\label{7}
\end{equation}
\\
If $\zeta \equiv \zeta _iX^i$ is a general element of dual space, then
\\
\begin{equation}
\label{8}
\left<\zeta ,\xi \right> = \zeta _i\xi ^i
\end{equation}
\\
The coadjoint action of $G$ in the dual space obeys
\\
\begin{equation}
\label{9}
\left<Ad^*_g\zeta ,Ad_g\xi \right> = \left<\zeta ,\xi \right> 
\end{equation}
\\
which yields
\\
\begin{equation}
\label{10}
Ad^*_g\zeta = \zeta_i'X^i,\qquad \zeta _i' = D^j\,_i(g^{-1})\zeta _j
\end{equation}
\\
On $\mathcal G^*$ one defines $Ad^*_g$-invariant Poisson bracket
\\
\begin{equation}
\label{11}
\left\{ \zeta _i,\zeta _j \right\} = c_{ij}\,^k\zeta _k
\end{equation}
\\
or, generally
\\
\begin{equation}
\label{12}
\left\{ f_1(\zeta ), f_2(\zeta )\right\} = \frac{\partial f_1}{\partial \zeta _i}\frac{\partial f_2}{\partial \zeta _j} c_{ij}\,^k \zeta _k
\end{equation}
\\
The above Poisson structure is degenerate; this is easily seen by considering the function $f(\zeta )$ corresponding to any Casimir operator
of $\mathcal G$.

Let us now introduce the dynamics by selecting a particular element of $\mathcal G$ as the Hamiltonian
\\
\begin{equation}
\label{13}
H = \alpha ^i X_i
\end{equation}
\\
Let, further,
\\
\begin{equation}
\label{14}
\xi (t) = Ad_{e^{-itH}} \xi = \xi ^i e^{-itH} X_i e^{itH} \equiv \xi ^i (t) X_i;
\end{equation}
\\
then one easily finds
\\
\begin{equation}
\label{15}
\dot \xi ^k(t) = c^k\,_{ij} \xi ^j(t)\alpha ^i
\end{equation}
\\
Defining the dynamics on dual space by
\\
\begin{equation}
\label{16}
\left< \zeta (t) , \xi (t) \right> = \left< \zeta (0) , \xi (0) \right>
\end{equation}
\\
we arrive at the following equations of motion 
\\
\begin{equation}
\label{17}
\dot \zeta _j(t) = c_{ji}\,^k \alpha ^i \zeta _k(t)
\end{equation}
\\
with the solution (cf. eqs.(\ref{10}) and (\ref{14})) 
\\
\begin{equation}
\label{18}
\zeta _j(t) = D^i\,_j\left( e^{itH}\right) \zeta _i(0)
\end{equation}
\\
Eqs.(\ref{17}) can be put in Hamiltonian form. Indeed, by defining 
\\
\begin{equation}
\label{19}
\tilde H (\zeta ) \equiv  \alpha ^i \zeta _i
\end{equation}
\\
we rewrite eq.(\ref{17}) as the canonical one
\\
\begin{equation}
\label{20}
\dot \zeta _i(t) = \left\{ \zeta _i(t) , \tilde H(\zeta (t))\right\}
\end{equation}
\\

Having defined the Hamiltonian dynamics in $\mathcal G ^*$ let us now analyze the action of $G$ as the symmetry
group. In doing that one should keep in mind that $H$, being an element of Lie algebra $\mathcal G$, does not commute
in general with symmetry generators. Therefore, they should depend explicitly on time.

Define the explicitly time dependent generators
\\
\begin{equation}
\label{21}
\tilde X_i \left( \zeta , t \right) \equiv D^j\,_i \left( e^{-itH} \right) \zeta _j
\end{equation}
\\
They obey the (Poisson) commutation rules of $\mathcal G$
\\
\begin{equation}
\label{22}
\left\{ \tilde X_i \left( \zeta , t \right) , \tilde X_j \left( \zeta , t \right) \right\} = c_{ij}\,^k \tilde X_k \left( \zeta , t \right)
\end{equation}
\\
Moreover, by virtue of eq.(\ref{18}) they are constants of motion
\\
\begin{equation}
\label{23}
\tilde X_i \left( \zeta (t) , t \right) = \zeta _i(0)
\end{equation}
\\
or
\\
\begin{equation}
\label{24}
\left\{ \tilde X_i \left( \zeta , t \right), \tilde H \left( \zeta \right) \right\} + 
\frac{\partial \tilde X_i \left( \zeta , t \right)}{ \partial t} = 0
\end{equation}
\\

The last equation can be also viewed as stating that $\tilde X_i \left( \zeta , t \right)$ generate symmetry transformation. Alternatively,
this is easily verified by considering infinitesimal transformations
\\
\begin{eqnarray}
\delta \zeta _i (t) &=& \left\{ \lambda ^k \tilde X_k \left( \zeta (t), t \right) , \zeta _i (t) \right\}
= \lambda ^k D^j\,_k \left( e^{-itH}\right) c_{ji}\,^l \zeta _l(t) =\nonumber\\
\label{25}
&=& \lambda ^k D^j\,_k \left( e^{-itH}\right) c_{ji}\,^l D^m\,_l \left( e^{itH} \right) \zeta _m(0) =
D^m\,_i \left( e^{itH} \right)\lambda ^k c_{km}\,^l \zeta _l (0) =\\
&=& D^m\,_i \left( e^{itH}\right) \left\{ \lambda ^k \zeta _k(0) , \zeta _m(0) \right\} =
D^m\,_i \left( e^{itH}\right) \delta \zeta _m(0)\nonumber
\end{eqnarray}
\\
The global counterpart of eq.(\ref{25}) defines the action of finite symmetry transformations at time $t$:
\\
\begin{equation}
\label{26}
\zeta '_i = D^m\,_i\left( e^{itH}\right) D^k\,_m\left(e^{-i\lambda ^l X_l}\right) D^j\,_k\left(e^{-itH}\right) \zeta _j(t)
\end{equation}
\\

We use above the exponential parametrization of global symmetry transformations; however, one has to keep in mind that
in most cases the exponential parametrization does not provide the global map for group manifold and the global topological 
properties must be carefully taken into account.  

In the Hamiltonian formalism the description of the symmetries refers to a fixed time. On the other hand, in Lagrangian 
(i.e. "explicitly covariant") approach one considers also the action of symmetry group on time variable. The necessity
of taking into account the nontrivial action of symmetry group on time variable appears often in the following context.
Let us consider the symmetry transformations acting in the phase space and pose the question whether they can be viewed
as canonical counterpart of some point transformations. In general, the answer is no because, for example, new coordinates
are not only the function of old ones but also depend on initial momenta. However, it can happen that the momentum dependence can be
accounted for by admitting the time variation. Then our initial canonical transformation is equivalent to the point one, provided
the latter includes time variation.

Let us assume that there exists a (generally nonlinear) realization of our symmetry group $G$ on one-dimensional manifold parametrized
by time. This is possible if $\mathcal G$ contains a subalgebra which after adjoining $H$ spans the whole $\mathcal G$. This means that
$G$ contains the subgroup $K$ such that $G/K$ is one-dimensional coset space generated by $H$. The action of $G$ on $t$ variable is given by
\\
\begin{equation}
\label{27}
t' = f(t;\underline \lambda ), \qquad f\left( f(t,\underline \lambda ');\lambda \right) = 
f\left(t;\varphi (\underline \lambda , \underline \lambda ')\right)
\end{equation}
\\
where $\varphi (\underline \lambda , \underline \lambda ')$ gives the composition law in some (in general - local) coordinates on $G$.
The full action of $G$  reads
\\
\begin{eqnarray}
t' &=& f(t;\underline \lambda )\nonumber\\
\zeta '_i(t') &=& D^m\,_i \left( e^{it'H}\right) D^k\,_m \left( g^{-1}\left(\underline \lambda \right) \right) D^j\,_k \left( e^{-itH} \right) \zeta _j(t)
\label{28}
\end{eqnarray}
\\

It is easy to verify that the action of $G$, defined above, obeys the appropriate composition rule; moreover, $\zeta '_i(t')$ obeys 
the equation of motion (\ref{17}) (with respect to the new time variable $t'$).\\
Note that the eqs. (\ref{28}) may be rewritten as follows. First, the second formula (\ref{28}) in a more compact form reads 
\\
\begin{equation}
\label{29}
\zeta '_i(t') = D^i\,_i \left( \left( e^{-it'H} g e^{itH}\right)^{-1}\right) \zeta _j(t)
\end{equation}
\\
Moreover, the action of $g$ on time variable is given, according to the remark above, by
\\
\begin{equation}
\label{30}
g e^{itH} = e^{it'H} k(g,t), \qquad k(g,t)\in K
\end{equation}
\\
which yields the first eq.(\ref{28}).

The dynamics considered up to now is rather trivial. In fact, eqs.(\ref{17}) are linear and their solution (\ref{18}) is immediately known.
However, as mentioned above, the Poisson bracket (\ref{11}) is degenerate so there exist in $\mathcal G^*$ submanifolds which are invariant 
under the dynamics (\ref{17}). The main point of the orbit method is that these submanifolds, under the assumption that $G$ acts transitively 
on them (i.e. the orbits of coadjoint action of $G$ on $\mathcal G^*$) provide all $G$-invariant phase manifolds with transitive action of $G$.\\
The Poisson bracket can be consistently restricted to the orbits and becomes then nondegenerate \cite{n6}-\cite{n11}.

The geometry of orbits may be described in terms of the geometry of $G$. Let us select a particular orbit $\mathcal O$ and let $\zeta ^{(0)} \in 
\mathcal O$ be any point on it. Let $\mathcal H \subset G$ be the stability subgroup of $\zeta ^{(0)}$:
\\
\begin{equation}
\label{31}
\zeta ^{(0)} = D^j\,_i\left( h^{-1}\right) \zeta _j^{(0)} \Longleftrightarrow h\in \mathcal H
\end{equation}
\\
Then $\mathcal O$ is isomorphic to the left coset space $ W = G/\mathcal H$. Using this isomorphism it is not difficult to give an explicit
expression for the symplectic  form on $\mathcal O$ defining the Poisson bracket (Kirillov form). Namely, let
\\
\begin{equation}
\label{32}
\omega (w) \equiv w^{-1}dw = i\omega ^i(w) X_i,\qquad w\in W
\end{equation}
\\
be the left-invariant Cartan form on $W$. Define
\\
\begin{equation}
\label{33}
\tilde \omega (w) = \omega ^k(w)\zeta _k^{(0)};
\end{equation}
\\
then $d\tilde \omega (w)$ is the Kirillov form on $\mathcal O$ \cite{n11}.\\
Due to the isomorphism of $W$ and $\mathcal O$ the coordinates $w$ on $W$ provide simultaneously the coordinates on $\mathcal O$. 
Therefore, we would like to rewrite the dynamical equations (\ref{17}) in terms of $w$. To this end we put
\\
\begin{equation}
\label{34}
\zeta _i(t) = D^k\,_i \left( w^{-1}(t)\right) \zeta _k^{(0)} = D^{-1}\left(w(t)\right)^k\,_i\zeta ^{(0)}_k
\end{equation}
\\
Inserting eq.(\ref{34}) into eq.(\ref{17}) one obtains
\\
\begin{equation}
\label{35}
\left( \Omega ^l\,_m(w,\frac{dw}{dt})D^{-1}(w)^m\,_i + c_{ij}\,^k \alpha ^j D^{-1}(w)^l\,_k \right) \zeta _l^{(0)} = 0, 
\end{equation}
\\
or
\\
\begin{equation}
\label{36}
\left( \Omega ^l\,_m(w,\frac{dw}{dt}) + c_{ij}\,^k \alpha ^j D(w)^i\,_m D^{-1}(w)^l\,_k \right) \zeta _l^{(0)} = 0
\end{equation}
\\
here $ \Omega ^l\,_m$ is the Cartan form $w^{-1}dw$ in adjoint representation
\\
\begin{equation}
\label{37}
\Omega ^l\,_m(w,\frac{dw}{dt})= \left( D^{-1}\right)^l\,_k(w)dD^k\,_m(w)
\end{equation}
\\

Note that due to the invariance of the tensors $c^i\,_{jk}$ under the adjoint representation eq.(\ref{36}) is equivalent to
\\
\begin{equation}
\label{38}
\left( \Omega ^l\,_m\left(w,\frac{dw}{dt}\right) + c^l\,_{mk}\left( D^{-1}\right) ^k\,_j(w)\alpha ^j\right) \zeta ^{(0)}_l =0
\end{equation}
\\

To solve the eq. (\ref{36}) note that the Hamiltonian in adjoint representation reads
\\
\begin{equation}
\label{39}
(iH)^i\,_j =\alpha ^kc^i\,_{jk}
\end{equation}
\\
This allows to rewrite eq. (\ref{38}) symbolically as
\\
\begin{equation}
\label{40}
\left( i  w^{-1}Hwdt + w^{-1}dw\right) \zeta ^{(0)} = 0\\
\end{equation}
\\
or
\\
\begin{equation}
\label{41}
\left(e^{itH}w(t)\right)^{-1} d\left(e^{itH}w(t)\right)\zeta ^{(0)} = 0
\end{equation}
\\
which implies that $\left(e^{itH}w(t)\right)^{-1}d\left(e^{itH}w(t)\right)$ belongs to the Lie algebra $\eta $ of $\mathcal H$. So we arrive 
at the following formula for equations of motion
\\
\begin{equation}
\label{42}
\left(e^{itH}w(t)\right)^{-1}d\left(e^{itH}w(t)\right) \in \eta 
\end{equation}
\\
which, in fact, yields the constraints on Cartan forms.
The general solution to (\ref{42}) reads
\\
\begin{equation}
\label{43}
e^{itH}w(t) = w_0h(t),\qquad w_0\in W,\qquad  h(t)\in \mathcal H
\end{equation}
\\
or
\\
\begin{equation}
\label{44}
e^{-itH}w_0 = w(t)h^{-1}(t)
\end{equation}
\\
We conclude that the dynamics is given by nonlinear action of $\exp{(-itH)}$ on coset space $G/\mathcal H$ (which is rather natural conclusion).

Consider now the action of the symmetry group $G$. By virtue of eqs.(\ref{29}) and (\ref{34}) one gets
\\
\begin{equation}
\label{45}
\zeta '_j(t') = D^k\,_j\left( \left(e^{-it'H}ge^{itH}w\right)^{-1}\right)\zeta ^{(0)}_k
\end{equation}
\\
In terms of geometry of coset space eq.(\ref{45}) yields
\\
\begin{equation}
\label{46}
e^{-it'H}ge^{itH}w = w'h(g,w,t), \qquad h(g,w,t) \in \mathcal H
\end{equation}
\\
or
\\
\begin{equation}
\label{47}
ge^{itH}w = e^{it'H}w'h(g,w,t)
\end{equation}
\\

We see that the action of symmetry group, as defined within the framework of the orbit method, coincides with the standard nonlinear 
realization of the symmetry group a la Coleman et al. \cite{n12}. In fact, let us compare eqs.(\ref{43}) and (\ref{47}). We see that,
starting from the actual state of the system $w(t)$ we first travel back to the initial moment, $t=0$, then act with the element $g$
of symmetry group and finally come back to the state at the transformed time $t'$
\\
\begin{equation}
\label{48}
ge^{itH}w(t) = gw_0h(t) =w'_0h(g,w_0)h(t) = e^{it'H}w'(t')h(t')
\end{equation}
\\
i.e. time-dependent action of $g$ consists in relating two states at $t$ and $t'$ which correspond to the two initial states related by
the standard action of $G$ (which, again, is not surprising).
\section{Examples}
\subsection{Conformal quantum mechanics}
\par
The conformal group in $1+0$-dimensions coincides with $SL(2,\mathbb R)$. It is generated by time translations $(H)$, dilatations $(D)$
and special conformal generator $(K)$. The corresponding Lie algebra reads
\\
\begin{equation}
\label{49}
\left[ D,H\right] = -iH,\quad \left[D,K\right] = iK, \quad \left[ K,H\right] = -2iD
\end{equation}
\\
Its basic representation is given by
\\
\begin{equation}
\label{50}
H = i\sigma _+,\qquad K = -i \sigma _-, \qquad D = -\frac{i}{2}\sigma _3
\end{equation}
\\
$SL(2,\mathbb R)$ is locally isomorphic to $SO(2,1)$. Indeed, defining
\\
\begin{eqnarray}
M_0 &=& \frac{1}{2}(H+K)\nonumber\\
\label{51}
M_1 &=& \frac{1}{2}(-H + K)\\
M_2 &=& D\nonumber
\end{eqnarray}
\\
one obtains
\\
\begin{equation}
\label{52}
\left[ M_\alpha ,M_\beta \right] = -i \epsilon _{\alpha \beta }\,^\gamma M_\gamma ,
\qquad \epsilon _{012} = 1,\quad g_{\alpha \beta } = diag(+--)
\end{equation}
\\
The counterpart of eq.(\ref{50}) is
\\
\begin{eqnarray}
\label{53}
M_0 &=& -\frac{1}{2} \sigma _2\nonumber\\
M_1 &=& -\frac{i}{2} \sigma _1\\
M_2 &=& -\frac{i}{2} \sigma _3\nonumber
\end{eqnarray}
\\
Due to the local isomorphism  of $SL(2,\mathbb R)$ and $SO(2,1)$ the adjoint (and, due to the semisimplicity, coadjoint) action
of $SL(2,\mathbb R)$ is the same as that of Lorentz group in $1+2$-dimensions. This allows for simple classification of orbits.
Let us take the orbit
\\
\begin{equation}
\label{54}
\left(\zeta _0\right)^2 - \left(\zeta _1\right)^2 - \left(\zeta _2\right)^2 = \lambda ^2,\qquad \zeta _0 > 0 
\end{equation}
\\
As the standard vector we take $\zeta^{(0)} = (\lambda ,0,0)$. Its stability subgroup is one-dimensional group generated by $M_0 = \frac{1}{2}(H+K)$.
The convenient coset parametrization reads
\\
\begin{equation}
\label{55}
w = e^{iw_1K}e^{iw_2D}
\end{equation}
\\
Let
\\
\begin{equation}
\label{56}
w_0 = e^{ic_1K}e^{ic_2D}
\end{equation}
\\
then eq. (\ref{44}) takes the form
\\
\begin{equation}
\label{57}
e^{itH}e^{iw_1K}e^{iw_2D}  = e^{ic_1K}e^{ic_2D}e^{-\frac{i\tau }{2}(H+K)}
\end{equation}
\\
which, up to notation and the choice of constants, coincides with eqs. (3.1) and (3.4) of Ref. \cite{n14}.Now, eq. (\ref{57}) implies 
$i(\omega _HH + \omega _KK + \omega _DD ) = -\frac{i}{2}\frac{d\tau }{dt}dt(H+K)$ i.e. $\omega _D=0$, $\omega _H = \omega _K$ on trajectories. \\
Eq. (\ref{57}) yields (after some redefinition of $\tau$ ):
\\
\begin{eqnarray}
w_1 &=& \left( \frac{e^{-c_2} + c^2_1 e^{c_2}}{2}\right) \sin \tau \nonumber\\
\label{58}
w_2 &=& -2\ln{\cos{\frac{\tau }{2}}} - \ln{(e^{-c_2} + c^2_1e^{c_2})}\\
t &=& \frac{e^{c_2}}{1 + c^2_1e^{2c_2}}\tan{(\frac{\tau }{2})} + \frac{c_1e^{2c_2}}{1 + c^2_1e^{2c_2}}\nonumber
\end{eqnarray}
\\

Using eq. (\ref{34}) together with (\ref{55}) we find the parametrization of our orbit in terms of variables $w_1$, $w_2$:
\\
\begin{eqnarray}
\zeta _0 &=& \lambda \left(\textrm{ch}w_2 + \frac{1}{2}w_1^2e^{w_2}\right)\nonumber\\
\label{59}
\zeta _1 &=& \lambda \left(\textrm{sh}w_2 - \frac{1}{2}w_1^2e^{w_2}\right)\\
\zeta _2 &=& -\lambda w_12e^{w_2}\nonumber
\end{eqnarray}
\\
The Poisson brackets
\\
\begin{equation}
\left\{ \zeta _\alpha ,\zeta _\beta \right\} = - \epsilon _{\alpha \beta }\,^\gamma \zeta _\gamma 
\label{60}
\end{equation}
\\
yield
\\
\begin{equation}
\label{61}
\left\{ w_1, w_2\right\} = -\frac{1}{\lambda }e^{-w_2}
\end{equation}
\\

The same result is obtained by computing the Cartan forms (\ref{32})
\\
\begin{equation}
\label{62}
\omega ^0 = e^{w_2}dw_1,\qquad \omega ^1 = e^{w_2}dw_1,\qquad \omega ^2 = dw_2
\end{equation}
\\
According to the eq. (\ref{33}) we find
\\
\begin{equation}
\label{63}
\tilde \omega = \lambda e^{w_2} dw_1
\end{equation}
\\
and
\\
\begin{equation}
\label{64}
d\tilde \omega = - \lambda e^{w_2}dw_1\wedge dw_2
\end{equation}
\\
in agreement with eq.(\ref{61}).

The Hamiltonian reads
\\
\begin{equation}
\label{65}
\tilde H = \lambda \left( w_1^2e^{w_2} + e^{-w_2}\right)
\end{equation}
\\
and implies the following equations of motion
\\
\begin{eqnarray}
\label{66}
\dot w_1 &=& -w_1^2 + e^{-2w_2}\nonumber\\
\dot w_2 &=& 2w_1
\end{eqnarray}
\\
again in agreement with eqs.(2.14) of Ref.\cite{n14}

Let us define the canonical variables
\begin{equation}
\label{67}
x = \sqrt{2\lambda }e^{\frac{w_2}{2}}, \qquad p = \sqrt{2\lambda }w_1 e^{\frac{w_2}{2}}
\end{equation}
\\
Then $x > 0$, $p \in \mathbb{R}$, $\{ x , p \} = 1$ and
\\
\begin{equation}
\label{68}
\tilde H(x,p) = \frac{p^2}{2} + \frac{2\lambda ^2}{x^2}
\end{equation}
\\
The action functional reads
\\
\begin{equation}
\label{69}
S =  \int \left( - \tilde \omega \omega  - \tilde H (w)dt \right) = \lambda \int \left( -e^{w_2}dw_1 - \tilde H (w) dt \right)
\end{equation}
\\
Up to an exact form $S$ can be rewritten as
\\
\begin{equation}
\label{70}
S = \int \left( w_1 e^{w_2}dw_2 - \tilde H (w) dt \right) = \int \left( p dx - \tilde H (x,p) dt \right)
\end{equation}
\\
which is the standard form of action for conformal mechanics.

It is also easy to find the action of $SL(2,\mathbb{R})$. To this end we consider eq. (\ref{46}) (or (\ref{47})). Taking into
account the definition of coset space (\ref{55}) we see that $h(g,w,t)=1$ and the action of $SL(2,\mathbb{R})$ is simply
given by group multiplication, again with perfect accordance with eq.(2.6) of Ref. \cite{n14}.

\subsection{Galilean conformal mechanics}

S. Fedoruk at al. \cite{n16} constructed, using the method of nonlinear realizations, the dynamical systems invariant under
the action of $N$-Galilean conformal symmetry with $N=2$. It is well known that the $N$-Galilean conformal algebras with $N$ even
do not admit (except in two dimension) the central extension. This makes the classification of coadjoint orbits more complicated
than in the case of centrally extended algebras with $N$-odd.

Fedoruk at al. construction is based on the following ingredients: (i) the choice of the subgroup $\mathcal H \subset G$ which defines
the coset space  $G/\mathcal H$; (ii) the construction of Cartan forms on $G/\mathcal H$; (iii) the application of the so-called 
inverse Higgs mechanism.

The advantage of this scheme is that one gets an algorithm which allows to produce various invariant dynamical equations. The difficulty
is that one cannot take for granted that the resulting equation have automatically Hamiltonian form. In fact, this depends 
on whether the selected subgroup $\mathcal H$ is or is not a stability group of some point on coadjoint orbit.

Let us remind the form of conformal algebra for $N=2$. It is spanned by $so(d)$ generators $J_{ij}$, $sl(2,\mathbb R)$ ones $K, D, H$
and additional generators $P_i$, $B_i$ and $F_i$ which span $D^{(1,1)}$ representation of $so(d) \oplus sd(2,\mathbb R)$ :
\\
\begin{eqnarray}
\label{71}
\left[ H , P_k \right ] = 0 \qquad \left [ H, F_k\right ] &= &2iB_k \quad \left[H,B_k\right] = iP_k\nonumber\\
\left[ K , P_k \right ] = -2iB_k \quad \left [ K, F_k\right ] &=& 0 \qquad \left[K,B_k\right] = -iF_k\\
\left[ D , P_k \right ] = - iP_k \quad\, \left [ D, F_k\right ] &=& iF_k \quad \, \left[ D,B_k\right] = 0\nonumber
\end{eqnarray}
\\
$P_k$, $B_k$ and $F_k$ themselves span abelian subalgebra.

We shall discuss two examples of coadjont orbits and the corresponding dynamical systems; for simplicity we assume $d=3$.
Defining the tensor $X_{\alpha i} $ by
\\
\begin{eqnarray}
\label{72}
X_{0i} &\equiv & \frac{1}{2}\left( P_i + F_i\right)\nonumber\\
X_{1i} &\equiv & \frac{1}{2}\left( P_i - F_i\right)\\
X_{2i} &\equiv & B_i\nonumber
\end{eqnarray}
\\
one finds
\\
\begin{eqnarray}
\label{73}
\left [ J_i, J_j\right] &= &i\epsilon _{ijk}J_k\nonumber\\
\left [ M_\alpha , M_\beta \right] &=& -i\epsilon _{\alpha \beta }\,^{\gamma }M_{\gamma }\nonumber\\
\left [ J_i, X_{\alpha j}\right] &=& i\epsilon _{ijk}X_{\alpha k}\\
\left [ M_\alpha , X_{\beta i}\right] &=& -i\epsilon _{\alpha \beta }\,^\gamma X_{\gamma i}\nonumber
\end{eqnarray}
\\
the remaining commutators being zero. With $J^i$, $M^\alpha $, $X^{\alpha i}$ being the dual basis, the general element
of dual space reads
\\
\begin{equation}
\label{74}
j_iJ^i + m_\alpha M^\alpha  + x_{\alpha i}X^{\alpha i}
\end{equation}

Assume first that the stability group contains $SO(3)$. The general element (\ref{74}) depends on four coefficients, $j_i$, $X_{\alpha i}$,
belonging to spin-1 irreducible representations of $so(3)$ and the scalars. Therefore, for $SO(3)$ to be contained in stability
subgroup of (\ref{74}) it is necessary that $j_i = 0$, $x_{\alpha i} = 0$. The stability group of such a point is generated by $J_i$ and $X_{\alpha i}$.
We are left with $SL(2,\mathbb R)$ group and the dynamics considered in previous subsection.

As a second example we choose initial point on the coadjoint orbit which breaks explicity $SO(d)$ symmetry. To this end let us first write out 
explicitly the coadjoint action using eqs. (\ref{3}) and (\ref{10}). We put
\\
\begin{equation}
\label{75}
g = e^{i\sum \omega ^\alpha M_\alpha }e^{i\sum \eta _i J_i}e^{i \sum y^\alpha \,_iX_{\alpha i}} 
\end{equation}
\\
which yields
\\
\begin{eqnarray}
\label{76}
g^{-1}X_{\beta j}g &=& \left(R^{-1}\right)_{kj}\left(\Lambda ^{-1}\right)^\alpha \,_{\beta }X_{\alpha k}\nonumber\\
g^{-1}J_j g &=& \left(R^{-1}\right)_{kj}\left(J_k - \epsilon _{kil}y^\alpha \,_i X_{\alpha l}\right)\\
g^{-1}M_{\alpha } g &=& \left(\Lambda ^{-1}\right)^\beta \,_{\alpha }\left(M_\beta   + 
\epsilon _{\beta \gamma }\,^\rho y^\gamma  \,_i X_{\rho i}\right)\nonumber
\end{eqnarray}
\\
where $R\in SO(3)$, $\Lambda \in SO(2,1)$. The orbit under consideration is selected by taking the initial point in the form:
\\
\begin{equation}
\label{77}
\underline \zeta  = \underline m_\alpha M^\alpha  + \underline j_k J^k + \underline x_{\alpha i}X^{\alpha i}
\end{equation}
\\
with $\underline j_k = 0$, $\underline m_\alpha = 0 $ , 
$\underline x_{\alpha i} = \underline \zeta _\alpha \cdot \underline s_i$ , $ \underline \zeta _\alpha  \equiv (\lambda ,0,0),\lambda >0$ , $\underline s_i \equiv  (0,0,1)$.\\
It is now easy to see that the stability subgroup is generated by $M_0$, $J_3$, $X_{03}$, $X_{11}$, $X_{12}$, $X_{21}$ and $X_{22}$.
We are then left with eight generators $M_1$, $M_2$, $J_1$, $J_2$, $X_{01}$, $X_{02}$, $X_{13}$ and $X_{23}$. 
Therefore the coset element can be written as
\\
\begin{equation}
\label{78}
w = e^{i\sum '\omega ^\alpha M_\alpha }e^{i\sum '\eta _i J_i}e^{i \sum 'y^\alpha \,_iX_{\alpha i}} 
\end{equation}
\\
where $\sum '\omega ^\alpha M_\alpha \equiv \omega ^1 M_1 + \omega ^2 M_2$, $ \sum '\eta _i J_i \equiv  \eta _1 J_1 + \eta _2 J_2$,
$\sum 'y^\alpha \,_iX_{\alpha i} \equiv y^0 \,_1X_{01} + y^0 \,_2X_{02} + y^1 \,_3X_{13} + y^2 \,_3X_{23}  \equiv \underline y^\alpha \,_iX_{\alpha i}$.\\
The resulting orbit reads finally
\\
\begin{eqnarray}
\label{79}
m_\alpha &=& \left( \Lambda ^{-1}\right) ^\beta \,_\alpha \left( \epsilon _{\beta \gamma }\,^\rho \underline y^\gamma \,_i
\underline\zeta _\rho \underline s_i \right)\nonumber\\
j_i &=& -\left( R^{-1}\right)_{kj} \epsilon _{kil}\underline y^\alpha \,_i \underline \zeta _\alpha \underline s_l\\
x_{\beta j} &=& \left( R^{-1}\right)_{kj} \left(\Lambda _{-1}\right)^\alpha \,_\beta \underline \zeta _\alpha \underline s_l\nonumber
\end{eqnarray}
\\
where $\Lambda $ (respectively $R$) is generated by the elements $M_1$, $M_2$ (respectively $J_1$, $J_2$). The orbit (\ref{75}) may be
rewritten in more elegant form by defining 
\\
\begin{eqnarray}
\label{80}
\zeta _\alpha &\equiv &\left( \Lambda ^{-1} \right) ^\beta \,_\alpha \underline \zeta _\beta\nonumber\\
s_k &\equiv & \left(R^{-1}\right)_{jk} \underline s_j\\
y^\alpha \,_i &\equiv & \left(\Lambda ^{-1}\right)_\beta \,^\alpha \left( R^{-1}\right)_{ki}\underline y^\beta \,_k\nonumber
\end{eqnarray}
\\
Then eqs (\ref{79}) take the form
\\
\begin{eqnarray}
\label{81}
m_\alpha &=& \epsilon _{\alpha \beta }\,^\gamma y^\beta \,_k \zeta _\gamma s_k\nonumber\\  
j_i &=& - \epsilon _{ijk}y^{\alpha }\,_j \zeta _\alpha  s_k\\
x_{\alpha i} &=& \zeta _\alpha s_i\nonumber
\end{eqnarray}
\\
The price one has to pay for the new form of orbit are constraints. First, from eqs. (\ref{80}) one gets
\\
\begin{equation}
\label{82}
s_is_i =1, \qquad \zeta _\alpha \zeta ^\alpha = \lambda ^2
\end{equation}
\\
Moreover, one has to reduce the number of independent components of $y^\alpha \,_i$. The relevant constraints read
\\
\begin{eqnarray}
\label{83}
\zeta _\alpha s_iy^\alpha \,_i &=& 0\nonumber\\
\epsilon _{\alpha \beta \gamma }\epsilon _{ijk}s_j\zeta ^\beta y^\gamma \,_k &=&0
\end{eqnarray}
\\
The constraints (\ref{83}) express the fact that there are only four independent variables $y^\alpha \,_i$. They can be replaced by
new ones
\\
\begin{eqnarray}
\label{84}
t_i & \equiv & y^\alpha \,_i\zeta _\alpha \nonumber\\
\eta ^\alpha & \equiv & y^\alpha \,_i s_i
\end{eqnarray}
obeying
\begin{eqnarray}
\label{85}
s_i t_i & = & 0\nonumber\\
\zeta _\alpha \eta ^\alpha &=& 0
\end{eqnarray}
\\
The orbit (\ref{81}) is now parametrized as
\\
\begin{eqnarray}
\label{86}
m_\alpha &=& \epsilon _{\alpha \beta }\,^\gamma \eta ^\beta \zeta _\gamma \nonumber\\
j_i &=& \epsilon _{ijk}s_jt_k\\
x_{\alpha i} &=& \zeta _\alpha s_i\nonumber
\end{eqnarray}
\\
with parameters being constarined by eqs. (\ref{82}) and (\ref{85}).
It is not difficult to find the relevant Poisson brackets for new basic variables:
\\
\begin{eqnarray}
\label{87}
\left\{ s_i, s_j \right\} &=& 0\nonumber\\
\left\{ s_i, t_k \right\} &=& \delta _{ik} - s_is_k\nonumber\\
\left\{ t_i, t_k \right\} &=& t_is_k - t_ks_i\nonumber\\
\left\{ \zeta _\alpha , \zeta _\beta \right\} &=& 0\\
\left\{ \zeta _\alpha , \eta ^\beta  \right\} &=& \delta _\alpha \,^\beta  - \frac{1}{\lambda ^2}\zeta _\alpha \zeta ^\beta \nonumber\\
\left\{ \eta ^\alpha , \eta^\beta   \right\} &=& \frac{1}{\lambda ^2} \left( \eta ^\alpha \zeta  ^\beta  - 
\eta ^\beta \zeta  ^\alpha \right ) \nonumber
\end{eqnarray}
\\
with all remaining Poisson brackets vanishing.\\
Finally, the Hamiltonian reads
\\
\begin{equation}
\label{88}
H = m_0 - m_1 = \left(\eta ^2\zeta _1 -\eta ^1\zeta _2 + \eta ^0\zeta _2 - \eta ^2\zeta _0\right)
\end{equation}
\\
Let us describe in some detail our model. The phase space is the product of cotangent bundles to the unit sphere and to the upper sheet of hyperboloid
$\zeta ^\alpha \zeta _\alpha = \lambda ^2$. The $SO(3) \times SL(2,\mathbb R)$ group acts naturally. On the other hand the action of the abelian
subgroup $\exp {(iz^\alpha \,_i X_{\alpha i})}$ reads
\\
\begin{eqnarray}
 s'_i &=& s_i\nonumber\\
 \zeta '^\alpha  &=& \zeta ^\alpha \nonumber\\
 \label{89}
 t'_i &=& t_i + (\delta _{ik} - s_is_k)\zeta_\alpha z^\alpha _k\\
 \eta '^\alpha & =& \eta ^\alpha + \left ( \delta ^\alpha _\beta  - \frac{1}{\lambda ^2}\zeta ^\alpha \zeta _\beta \right ) z^\beta _is_i\nonumber
 \end{eqnarray}
 \\
 One can easily check that the Poisson structure (\ref{87}) is invariant under the above action.
 
 The dynamics of $s_i$ and $t_i$ variables is trivial. On the other hand the remaining equations of motion read
 \\
 \begin{eqnarray}
 \dot \zeta _0 &=& \zeta _2 - \frac{2}{\lambda ^2}(\zeta _0)^2\zeta _2\nonumber\\
 \dot \zeta _1 &=& -\zeta _2 - \frac{2}{\lambda ^2}\zeta _0\zeta _1\zeta _2\nonumber\\
 \dot \zeta _2 &=& -\zeta _0 + \zeta _1 - \frac{2}{\lambda ^2}\zeta _0 (\zeta _2)^2\nonumber\\
 \label{90}
 \dot \eta  _0 &=& \eta ^2 + \frac{2}{\lambda ^2}\eta ^0\zeta _0\zeta _2\\
 \dot \eta  _1 &=& -\eta ^2 + \frac{2}{\lambda ^2}\eta ^1\zeta _0\zeta _2\nonumber\\
 \dot \eta  _2 &=& -\eta ^0 + \eta ^1 + \frac{2}{\lambda ^2}\eta ^2\zeta _0\zeta _2\nonumber
\end{eqnarray}
\\
They are consistent with the constraints (\ref{82}), (\ref{85}).

One can easily construct the explicitly time dependent symmetry generators and verify that they obey the correct Poisson algebra of our conformal
Galilei group.

It is also straightforward but slightly tedious to show that the orbit construction can be reformulated in terms of the geometry
of nonlinear realizations defined by the appropriate choice of stability subgroup (generated by $J_3$, $M_0$, $X_{03}$, $X_{11}$, $X_{12}$, $X_{21}$ 
and $X_{22}$). The relevant constraints on Cartan forms are read off from eqs. (\ref{42}) and (\ref{43}). We omit the detailes here. 

As it has been mentioned above the full classification of coadjoint orbits for $N$-conformal Galilei algebras with $N$ even 
(i.e. not admitting the central extension) is rather involved even for $N=2$. For the special choice of the orbit, eq. (\ref{77}), the relevant
dynamical system is described by eqs. (\ref{81}) - (\ref{90}). However, the particular form of dynamics seems to depend strongly on the choice 
of the orbit. This is also the case for $N$ odd if we assume the central charge is vanishing. On the contrary, for nonvanishing central charge
the dynamics is essentially unique \cite{n18}, \cite{n19}.
\section{Conclusions}

There are two basic methods of explicit construction of dynamical systems on which a given group $G$ acts transitively as the symmetry group. One
is based on the technique of nonlinear realizations \cite{n12}. It allows for elegant and algorithmic construction of invariant dynamical equations 
by selecting an appropriate coset space and computing the relevant Cartan-Maurer forms (and applying the so-called inverse Higgs phenomenon \cite{n16}).
The main problem with this approach is that it does not always lead to the dynamics which admits Hamiltonian form.
\\
The second method is based on the idea of coadjoint orbits. They are equipped with invariant symplecting form (Kirillov form) and exhaust the list
of all symplectic manifolds on which a given group acts transitively as a group of canonical transformations. It appears to be the dynamical symmetry group provided
the Hamiltonian belongs to its Lie algebra (or universal enveloping algebra). A coadjoint orbit can be identified with an appropriate coset space
once the stability subgroup of an arbitrarily selected point of the orbit is determined. This identification allows to establish the relation between
both methods; in particular, the Kirillov form is expressible in terms of Cartan forms \cite{n11}. Hamiltonian equations can be also rewritten 
with the help of these forms. These general considerations were illustrated by few examples. First, we showed that the elegant result of Ivanov at al. 
concerning the conformal (i.e. $SL(2,\mathbb R)$) invariant mechanics fits perfectly into the framework of orbit method. The second example is
related to the $N=2$ conformal Galilei group. We find that if the stability subgroup contains $SO(3)$ then the Hamiltonian dynamics reduces to that
of $SL(2,\mathbb R)$ conformal mechanics. More complicated example has been also constructed which describes the system with four degrees of freedom
running over nontrivial configuration space $ S^2 \times \mathcal H^2$, where $\mathcal H^2$ is the upper sheet of the hyperboloid 
$\zeta _0^2 - \zeta _1^2 - \zeta _2^2 = \lambda ^2$.

\section{Acknowledgments}

The author would like to thank Professors Piotr Kosiński, Paweł Maślanka and Dr. Krzysztof Andrzejewski for helpful discussions and useful remarks.\\
This work is supported in part by MNiSzW Grant No.N202331139.

\end{document}